\documentclass{aa}  
\usepackage{graphicx}
\usepackage[english]{babel}
\usepackage{txfonts}
\newcommand{\text}{\mbox}
\newcommand{\vi}{\mathsf{V}_I}
\newcommand{\vq}{\mathsf{V}_Q}
\newcommand{\vu}{\mathsf{V}_U}
\newcommand{\vv}{\mathsf{V}_V}

\renewcommand{\dagger}{\star}

\begin{document}
   \title{Sensitivity of a bolometric interferometer to the CMB power spectrum}

   \subtitle{}

   \author{J.-Ch. Hamilton\inst{1} \and R. Charlassier\inst{1} \and C. Cressiot\inst{1} \and  J. Kaplan\inst{1} \and M. Piat\inst{1} \and C. Rosset\inst{2}}

   \offprints{\tt hamilton@apc.univ-paris7.fr}

   \institute{APC, Universit\'e Denis Diderot-Paris 7, CNRS/IN2P3, CEA, Observatoire de Paris ; 10 rue A. Domon \& L. Duquet, Paris, France
   \and
   	LAL, Laboratoire de l'acc\'el\'erateur Lin\'eaire, Universit\'e Paris-Sud 11, CNRS/IN2P3 ; B\^atiment 200, 91898 Orsay Cedex, France}

   \date{Received ; accepted }

 
  \abstract
   {The search for B-mode polarization fluctuations in the Cosmic Microwave Background is one of the main challenges of modern cosmology. The expected level of the B-mode signal is very low and therefore requires the development of highly sensitive instruments with low systematic errors. An appealing possibility is bolometric interferometry.}
   {We compare in this article the sensitivity on the CMB angular power spectrum achieved with direct imaging, heterodyne and bolometric interferometry.}
   {Using a simple power spectrum estimator, we calculate its variance leading to the counterpart for bolometric interferometry of the well known Knox formula for direct imaging.}
   {We find that bolometric interferometry is less sensitive than direct imaging. However, as expected, it is finally more sensitive than heterodyne interferometry due to the low noise of the bolometers. It therefore appears as an alternative to direct imagers with different and possibly lower systematic errors, mainly due to the absence of an optical setup in front of the horns.}
   {}

   \keywords{Cosmology -- Cosmic Microwave Background -- Inflation - Bolometric Interferometry}

   \maketitle
%

\section*{Introduction}
The detection of primordial gravity waves through B-mode polarization anisotropies in the Cosmic Microwave Backgroud is one of the most exciting challenges of modern cosmology. It could provide direct information on the energy scale of inflation, possibly associated with GUT~(\cite{liddle}). It would also allow one to investigate the standard cosmological model in detail through consistency tests involving the spectral indices of scalar and tensor perturbations and their amplitude ratio~(\cite{langlois}).

Despite the weakness of the expected signal, many teams have decided to join the quest for the B-modes  and to construct dedicated instruments that must combine exquisite sensitivity and precise control of systematic effects. Most of the projects proposed up to now use direct imagers, a concept that has proven to be very sensitive. However they might be affected by significant systematic effects such as ground-pickup and beam differences that would less affect an interferometer having no optics before the entry horns. Thus, we investigate the possibility of developing a high sensitivity interferometer dedicated to B-mode searches. A bolometric interferometer would combine the high sensitivity of bolometers with the clean optics of an interferometer and could complement the ongoing imaging projects.

In this article we investigate the sensitivity achieved by such an instrument and compare it with direct imagers and heterodyne interferometers. In section~\ref{defs} we review useful quantities regarding interferometry in general and the reconstruction of visibilities with a bolometric interferometer. We define in section~\ref{powspec} a simple power spectrum estimator under the assumption that $E$ and $B$ visibilities can be extracted from $Q$ and $U$ Stokes parameter visibilities (this is in itself an important issue and is beyond the scope of this article). From this estimator we derive a formula for the $C_\ell$ uncertainty with a bolometric interferometer equivalent to the famous one derived in~(\cite{knox}) for imagers. We compare the sensitivities achieved with imagers and bolometric and heterodyne interferometers in section~\ref{comp}. We discuss our results in section~\ref{discussion}.

\section{Definitions and assumptions\label{defs}}
\subsection{Useful interferometric quantities}
Here, we discuss basic definitions regarding interferometry in general and bolometric interferometry in particular. An interferometer observes the {\em visibilities} of an incoming radiation field $S(\vec{n})$ that are defined as:
\begin{equation}
\mathsf{v}_S(\vec{u})=\int S(\vec{n}) A(\vec{n})\exp(2i\pi~\vec{u}\cdot\vec{n})\mathrm{d}\vec{n}
\end{equation}
where $\vec{u}$ is the baseline defined as the vector separation between the horns $\vec{D}$ in units of the electromagnetic wavelength of the radiation $\vec{u}=\vec{D}/\lambda$.  The way visibilities are actually reconstructed in a bolometric interferometer is described in section~\ref{visrec}.
The beam of the input horns $A(\vec{n})$ is normalized to one at maximum. In the flat-sky approximation, one can write the visibility as a convolution of the Fourier modes of the incoming radiation with the Fourier transform of the input beam:
\begin{equation}
\mathsf{v}_S(\vec{u})=\int  \tilde{S}(\vec{u}) \tilde{A}(\vec{u}-\vec{v})\mathrm{d}\vec{v}= \tilde{S}(\vec{u})\otimes \tilde{A}(\vec{u})
\end{equation}
where the $\tilde{~}$ denotes the Fourier transform. Note that outside the flat-sky approximation, a similar formula is to be expected, although more complicated~(\cite{bunnwhite}). This convolution expression states that the sky cut in real space performed by the input beam is expressed as a convolution in Fourier space. The correspondance with multipoles is $\ell=2\pi \mathrm{u}$.
Each horn covers a solid angle $\Omega$ defined as:
\begin{eqnarray}
\Omega=\int A(\vec{n})\mathrm{d}\vec{n}.
\end{eqnarray}
If we approximate the beam by a Gaussian defined by its RMS $\sigma$, its FWHM is $\sigma\times2\sqrt{2\mathrm{ln}2}\simeq 2.35 \sigma$
and in that case the solid angle subtended by the horn is $\Omega=2\pi\sigma^2$
which can be related to the fraction of the sky observed with the horn $f_\mathrm{sky}=\Omega/4\pi$.
We assume that the horns are placed on a square grid so that the baselines are also located on a square grid. The minimum spacing between two horns is obtained by packing at a distance equal to their diameter. As they are diffraction limited, their section, solid angle and wavelength are such that\footnote{Actually, for a Gaussian beam, the illumination function on the entry of the horn is not flat (but is Gaussian) and the size of the horn has to be larger that what is quoted here by a factor of about 2.} $S\Omega\simeq \lambda^2$ so that their distance is $D_h=2\lambda/\sqrt{\pi\Omega}$. The spacing between visibilities in Fourier space is therefore:
\begin{eqnarray}
\mathrm{u}_\mathrm{min} &=& \frac{2}{\sqrt{\pi \Omega}}=\frac{1}{\pi \sqrt{f_\mathrm{sky}}}.
\end{eqnarray}
In our Gaussian approximation, the Fourier transform of the primary beam is:
\begin{eqnarray}
\tilde{A}(\vec{u})&=& \Omega \exp\left(-\pi\Omega\vec{u}^2\right)
\end{eqnarray}
so that in terms of a Gaussian in Fourier space, the resolution is:
\begin{equation}
\sigma_\mathrm{u}=\frac{1}{\sqrt{2\pi\Omega}}.
\end{equation}
The baselines are separated by $\mathrm{u_{min}}$ and each point in baseline space has a resolution $\sigma_\mathrm{u}$, therefore:
\begin{eqnarray}
\frac{\mathrm{u_{min}}}{\sigma_\mathrm{u}} &=& 2\sqrt{2}\simeq 3.
\end{eqnarray}
We can conclude that {\bf the different baselines are almost independent}.
In reality, the distribution of the electric field vanishes outside the horn aperture so that the primary beam is not an exact Gaussian and the beam in Fourier space is also truncated at a radius of $\mathrm{u}_\mathrm{min}$~(\cite{white}). This does not change the fact that different baselines are almost independent.


\subsection{Visibility reconstruction in bolometric interferometry\label{visrec}}
In a heterodyne interferometer, the visibilities are directly obtained using a correlator that gives as an output the correlated signal $E_1 E^\star_2$ coming from two antennae in a coherent way. Bolometers are incoherent detectors that measure the time averaged incoming power. Bolometric interferometers are therefore additive interferometers where the visibilities are obtained from squaring the sum of the signal coming from two horns : $P=\left< \left| E_1+E_2\right|^2\right>= \left< \left| E_1\right|^2\right>+\left< \left| E_2\right|^2\right>+2 \left< E_1E_2^\star \right>$. When a large number of horns are used, the signal detected by the bolometers is a linear combination of all available visibilities. The use of phase shifters on each of the input channels allows one to reconstruct the complex visibilities of the four Stokes parameters $\vi$, $\vq$, $\vu$ and $\vv$. We have shown~(\cite{coherentsummation}) that this reconstruction is done optimally when the phase-shifting scheme is such that equivalent baselines\footnote{{\em Equivalent baselines} are sometimes called {\em redundant baselines} in the literature. They correspond to different pairs of horns separated by the same vector and therefore corresponding to the same point in $(u,v)$ space.} are summed coherently -- they correspond to the same phase difference. In this case, the noise covariance matrix on the complex reconstructed visibilities is diagonal and has the form:
\begin{eqnarray}
\mathcal{N}_{ij}&=& \delta_{ij}\frac{4~\mathrm{NET}^2 \Omega^2 N_h}{N_t}\frac{1}{N_\mathrm{eq}^2(i)}.
\end{eqnarray}
The first factor $2$ comes from the fact that when measuring polarized visibilities, one cannot have access to both $Q$ and $U$ Stokes parameters at the same time~(\cite{coherentsummation}). The second is due to the fact that here we are dealing with the covariance matrix of the complex visibilities instead of their real or imaginary parts as in~(\cite{coherentsummation}). $N_h$ is the number of entry horns, $N_t$ is the number of time samples, $N_\mathrm{eq}(i)$ is the number of equivalent baselines corresponding to baseline $i$ and NET is the noise equivalent temperature expressed in $\mathrm{\mu K/\sqrt{Hz}}$. An extensive analytical and Monte-Carlo based study of the reconstruction of the visibilities in bolometric interferometry can be found in~(\cite{coherentsummation}).

\section{Power spectrum estimator\label{powspec}}
\subsection{E and B fields from the Stokes parameters}
In the flat-sky approximation, the $E$ and $B$ polarization fields are related to the Stokes parameters by a simple rotation of angle $\phi$, the angle between $\vec{u}$ and the $u_x$ axis~(\cite{zald}). In terms of visibilities, this can be written:
\begin{equation}
\left\{\begin{array}{lll}
\vq(\vec{u})&=&\int \left[ \cos2\phi~ \tilde{E}(\vec{v})-\sin2\phi~\tilde{B}(\vec{v})\right] \tilde{A}(\vec{u}-\vec{v})\mathrm{d}\vec{v} \\
\vu(\vec{u})&=&\int \left[ \sin2\phi ~\tilde{E}(\vec{v})+\cos2\phi~\tilde{B}(\vec{v}) \right] \tilde{A}(\vec{u}-\vec{v})\mathrm{d}\vec{v} \end{array}\right.
\end{equation}
For the simplicity, we do not discuss the $E$/$B$ separation here and assume that one can obtain a set of pure $E$ and $B$ visibilities  from the Stokes parameter visibilities defined as\footnote{Obtaining such pure $E$ and $B$ modes is a complex issue in itself and deserves a full study.}:
\begin{equation}
\left\{\begin{array}{lll}
\mathsf{V}_E(\vec{u})&=&\int  \tilde{E}(\vec{v})\tilde{A}(\vec{u}-\vec{v})\mathrm{d}\vec{v} \\
\mathsf{V}_B(\vec{u})&=&\int\tilde{B}(\vec{v})  \tilde{A}(\vec{u}-\vec{v})\mathrm{d}\vec{v} \end{array}\right.
\end{equation}
and the covariance matrix of the $B$ visibilities contains the $BB$ angular power spectrum:
\begin{eqnarray}
\left< \mathsf{V}_B(\vec{u})\mathsf{V}_B^\star(\vec{u}')\right>&=& \int \left< \tilde{B}(\vec{v})\tilde{B}^\star(\vec{v}')\right> \tilde{A}(\vec{u}-\vec{v})\tilde{A}^\star(\vec{u}'-\vec{v}')\mathrm{d}\vec{v}\mathrm{d}\vec{v}' \\
&=& \int C_\ell^{BB}(\vec{v})\tilde{A}(\vec{u}-\vec{v})\tilde{A}^\star(\vec{u}'-\vec{v})\mathrm{d}\vec{v}\\
&=& \delta(\vec{u}-\vec{u}')\times  \int C_\ell^{BB}(\vec{v}) \left| \tilde{A}(\vec{u}-\vec{v})\right|^2\mathrm{d}\vec{v}
\end{eqnarray}
the last equation comes from the fact that the different baselines we measure are independent from the beam point of view.

In the presence of noise and assuming the power to be flat enough to be taken out of the integral (recall that $\ell=2\pi\mathrm{u}$):
\begin{equation}
\left< \mathsf{V}_B(\vec{u})\mathsf{V}_B^\star(\vec{u}')\right>= \delta(\vec{u}-\vec{u}')\times   C_\ell^{BB} \underbrace{\int \left| \tilde{A}(\vec{v})\right|^2\mathrm{d}\vec{v}}_{=\Omega/2} +\mathcal{N}(\vec{u},\vec{u}').
\end{equation}
As said before, the noise covariance matrix of the Stokes parameter visibilities is diagonal and we assume that it is still the case for that of $\mathsf{V}_E$ and $\mathsf{V}_B$. Labeling $\vec{u}$ and $\vec{u}'$ with indices $i$ and $j$ and $\mathsf{V}_B(\vec{u})$ and $\mathsf{V}_B(\vec{u}')$ by $\mathsf{V}_i$ and $\mathsf{V}_j$, one gets:
\begin{equation}
\left< \mathsf{V}_i \mathsf{V}^\star_{j}\right> =C_\ell \frac{\Omega}{2}\delta_{ij} +\mathcal{N}_{ij}\delta_{ij}.
\end{equation}

\subsection{A simple pseudo-power spectrum  estimator and its variance}
From the above equation, it is obvious that the simplest unbiased estimator of the power spectrum is:
\begin{equation}
C_\ell = \frac{2}{\Omega} \times \frac{1}{N_\neq(\ell)} \sum_{i=0}^{N_\neq(\ell)-1} (\mathsf{V}_i\mathsf{V}^\star_i-\mathcal{N}_{ii})
\end{equation}
where $N_\neq(\ell)$ is the number of different baselines corresponding to multipole $\ell$.
The variance of this estimator is ($\mathsf{C}_\ell$ is the {\bf true} power spectrum):
\begin{eqnarray}
\mathrm{Var}(C_\ell)&=& \left< C_\ell^2\right> -\mathsf{C}_\ell^2 \\
&=& \left(\frac{2}{\Omega N_\neq(\ell)}\right)^2 \left< \left[ \sum_i (\mathsf{V}_i\mathsf{V}^\star_i-\mathcal{N}_{ii})\right]^2\right> -\mathsf{C}_\ell^2 \\
&=& \frac{1}{N^2_\neq(\ell)}\sum_i \left( \mathsf{C}_\ell + \frac{2}{\Omega}\mathcal{N}_{ij}\right)^2
\end{eqnarray}
where we used Wick's theorem to calculate the fourth order moments and the fact that each of the $N_\neq(\ell)$ different baselines contributing to $C_\ell$ is measured independantly so that their variances add linearly. If one makes the additional assumption that all of these different baselines have the same noise variance $\mathcal{N}_{ij}=\sigma_\mathsf{V}^2~\delta_{ij}$, the error on the power spectrum reads:
\begin{equation}
\Delta C_\ell = \sqrt{\frac{1}{N_\neq(\ell)}}\left( \mathsf{C}_\ell +\frac{2\sigma_\mathsf{V}^2}{\Omega}\right)
\end{equation}
which is the equivalent for interferometry of the well known imaging-oriented Knox formula~(\cite{knox}). The expression for $N_\neq(\ell)$ is the number of different modes one can have access to at a given $\ell$. We assume that we are considering a bin in visibility space $\Delta\mathrm{u}=\Delta\ell /2\pi$ centered at $\mathrm{u}$; the number of modes is the ratio between the available surface of the bin $\pi \mathrm{u}\Delta\mathrm{u}$ (we only consider the top part of the Fourier plane as the modes in the bottom part are the same) to the effective surface of the beam in Fourier space $2\pi \sigma_\mathrm{u}^2$ :
\begin{eqnarray}
N_\neq(\ell)&=&\frac{\pi \mathrm{u}\Delta\mathrm{u}}{2\pi \sigma_\mathrm{u}^2} = \ell\Delta\ell f_\mathrm{sky}
\end{eqnarray}
we therefore find the same formula as for an imager, except for the noise part of course:
\begin{equation}\label{eqerrcl}
\Delta C_\ell^\mathrm{BI} = \sqrt{\frac{2}{2\ell f_\mathrm{sky} \Delta\ell}}\left( \mathsf{C}_\ell +\frac{2\sigma_\mathsf{V}^2}{\Omega}\right)
\end{equation}
where the noise on the visibilities is taken from~(\cite{coherentsummation}).
\begin{equation}
\sigma_\mathsf{V}^\mathrm{BI}=\sqrt{\frac{N_h}{N_\mathrm{eq}}} \times \frac{2~\mathrm{NET_\mathrm{BI}}~\Omega}{\sqrt{N_\mathrm{eq}}\sqrt{N_t}}.
\end{equation}

\section{Comparison with an imager and a heterodyne interferometer\label{comp}}
\subsection{Analytical formulae}
The above expression is the same for both heterodyne interferometry and direct imaging, only the expression of $\sigma_\mathsf{V}$ changes. 
For heterodyne interferometry, if the noise equivalent temperature of one of the two input channels of the correlator is $\mathrm{NET_{HI}}$, the noise on the reconstructed Stokes parameter visibility calculated with $N_t$ time samples and averaged over $N_\mathrm{eq}$ equivalent baselines is given by:
\begin{equation}
\sigma_\mathsf{V}^{HI}= \frac{2\sqrt{2}~\mathrm{NET_\mathrm{HI}}~\Omega}{\sqrt{N_\mathrm{eq}}\sqrt{N_t}}.
\end{equation}
The first factor 2 comes from the multiplication of the two sine waves, a factor $\sqrt{2}$ from the fact that two correlators are involved when calculating Stokes parameters visibilities, another  factor $\sqrt{2}$ appears because we are talking about the noise on the complex visibility instead of its real or imaginary part. Finally a factor $1/\sqrt{2}$ is regained because two sets of independent measurements of the Stokes parameters visibilities can be simultaneously obtained if one forms all the possible complex correlations.
The expression we find is in agreement with~(\cite{hobson}) and~(\cite{white}).
In the direct imaging case, the error on the power spectrum is taken from~(\cite{knox}) and adapted to partial sky polarized measurements:
\begin{equation}
\Delta C_\ell^\mathrm{Im} = \sqrt{\frac{2}{(2\ell+1) f_\mathrm{sky} \Delta\ell}}\left( \mathsf{C}_\ell +\frac{4\mathrm{NET_{Im}^2}\Omega}{N_h B_\ell^2 N_t}\right)
\end{equation}
where $B_\ell=\exp(-\ell^2\sigma_\mathrm{beam}^2/2)$ is the imager's beam transfer function.
In the imaging case $\Omega$ is of course defined as the solid angle covered on the sky $\Omega=4\pi f_\mathrm{sky}$ (the fact that  the integral of the primary beam is the total solid angle covered on the sky is specific to interferometry). Note that the factor of 4 is obtained by a factor of 2 on the polarized NET for polarization sensitive bolometers and another factor of 2 due, as before, to the fact that $Q$ and $U$ cannot be obtained at the same time.

As the sample variance term is exactly the same whatever technique is used (as expected), we are only interested in comparing the noise terms. We assume in the following that we are comparing three instruments observing the same fraction of the sky $f_\mathrm{sky}$ from the ground for the same duration:
\begin{itemize}
\item a direct imager with $N_h$ horns, an angular resolution given by $\sigma_\mathrm{beam}$ and a  $\mathrm{NET}_\mathrm{Im}\simeq 150~\mathrm{\mu K}/\sqrt{\mathrm{Hz}}$, as stated for Clover in~(\cite{clover2}) at 97GHz in Chile.
\item a heterodyne interferometer with a primary beam covering $f_\mathrm{sky}$, using a square array of $N_h$ input channels each with $\mathrm{NET}_\mathrm{HI}\simeq 250~\mathrm{\mu K}/\sqrt{\mathrm{Hz}}$, as stated for QUIET in~(\cite{quiet}) at 90 GHz in Chile.
\item a bolometric interferometer with the same characteristics as the heterodyne one but with a NET identical to that of a bolometric imager $\mathrm{NET}_\mathrm{BI}\simeq 150~\mathrm{\mu K}/\sqrt{\mathrm{Hz}}$.
\end{itemize}
The choice of 90GHz is motivated by the fact  that a packed array of $~20$ degrees FWHM primary horns interferometer operating at these frequencies would cover the multipole range relevant for primordial B-mode signals ($25<\ell<200$). It is also at these frequencies that coherent and bolometric detectors can operate simultaneously.
We will use the direct imager as a reference and calculate the ratio of the direct imager's noise error to that of the interferometers. This ratio therefore should be less than one if the direct imager is more sensitive from the strict noise point of view.\\
For the bolometric interferometer, one gets:
\begin{eqnarray}
\left. \frac{\Delta C_\ell^\mathrm{Im}}{\Delta C_\ell^\mathrm{BI}} \right|_\mathrm{noise}&=& \frac{1}{2}\times \left(\frac{N_\mathrm{eq}}{N_h}\right)^2 \times\frac{1}{B_\ell^2} \times \left(\frac{\mathrm{NET}_\mathrm{Im}}{\mathrm{NET}_\mathrm{BI}}\right)^2\\
&=& \frac{1}{2}\times \left(\frac{N_\mathrm{eq}}{N_h}\right)^2 \times\frac{1}{B_\ell^2}
\end{eqnarray}
as the $\mathrm{NET}$ are the same for the bolometers used for imaging or for bolometric interferometry. \\
For heterodyne interferometry, one gets:
\begin{eqnarray}
\left. \frac{\Delta C_\ell^\mathrm{Im}}{\Delta C_\ell^\mathrm{HI}}\right|_\mathrm{noise} &=&  \frac{1}{4}\times \left(\frac{N_\mathrm{eq}}{N_h}\right) \times\frac{1}{B_\ell^2}\times \left(\frac{\mathrm{NET}_\mathrm{Im}}{\mathrm{NET}_\mathrm{HI}}\right)^2
\end{eqnarray}
We can see that the ratio $N_\mathrm{eq}/N_h$ is always less than one, which gives a clear advantage to direct imaging from the strict point of view of the noise. This ratio appears squared in the ratio of imaging to bolometric interferferometry and without power in the ratio of imaging to heterodyne interferometry but in the latter case, the NET ratio is also less  than one,  penalising heterodyne interferometry.
\subsection{Approximate expressions and simulations}
\begin{figure*}[!t]
\centering\resizebox{\hsize}{!}{\centering{
\includegraphics{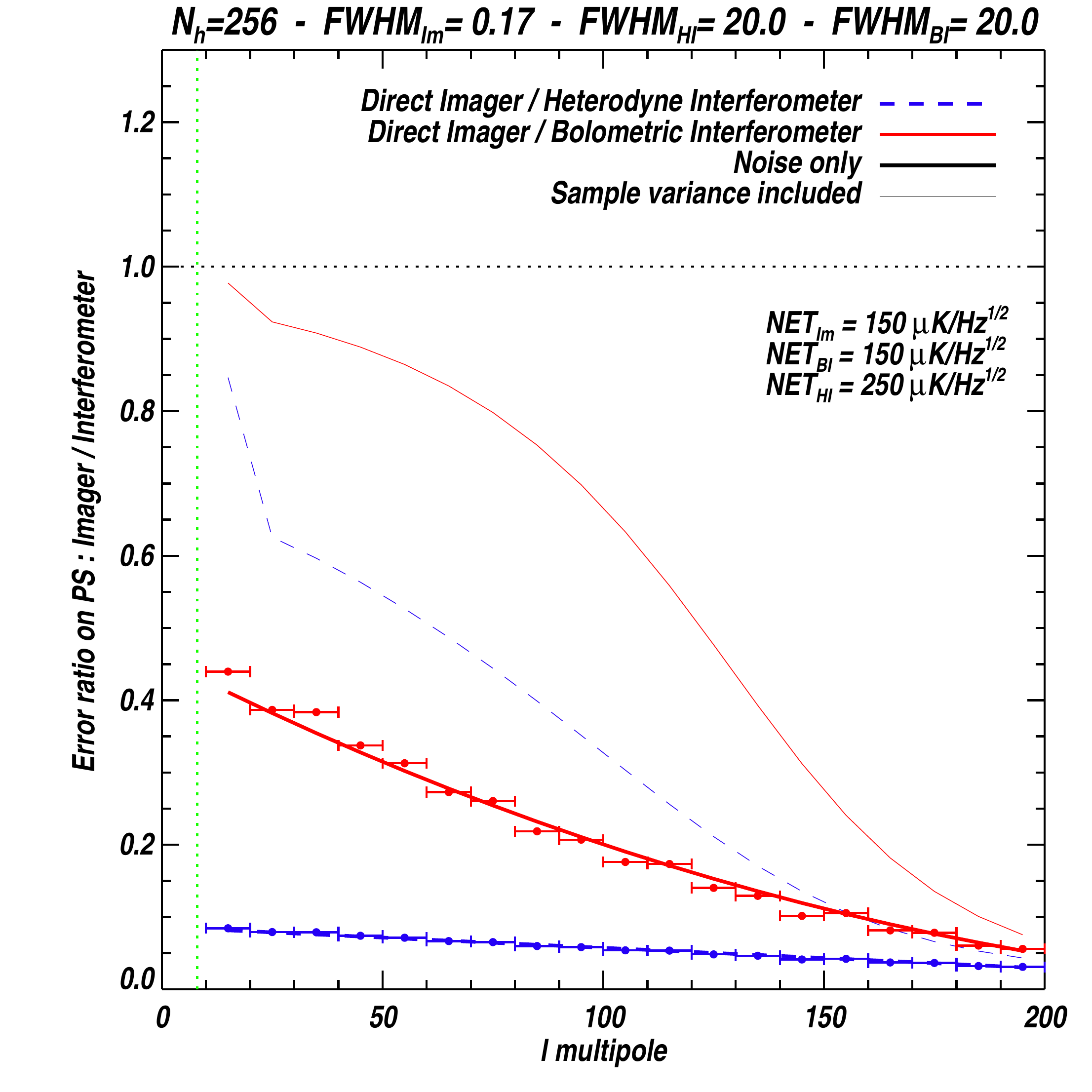}\includegraphics{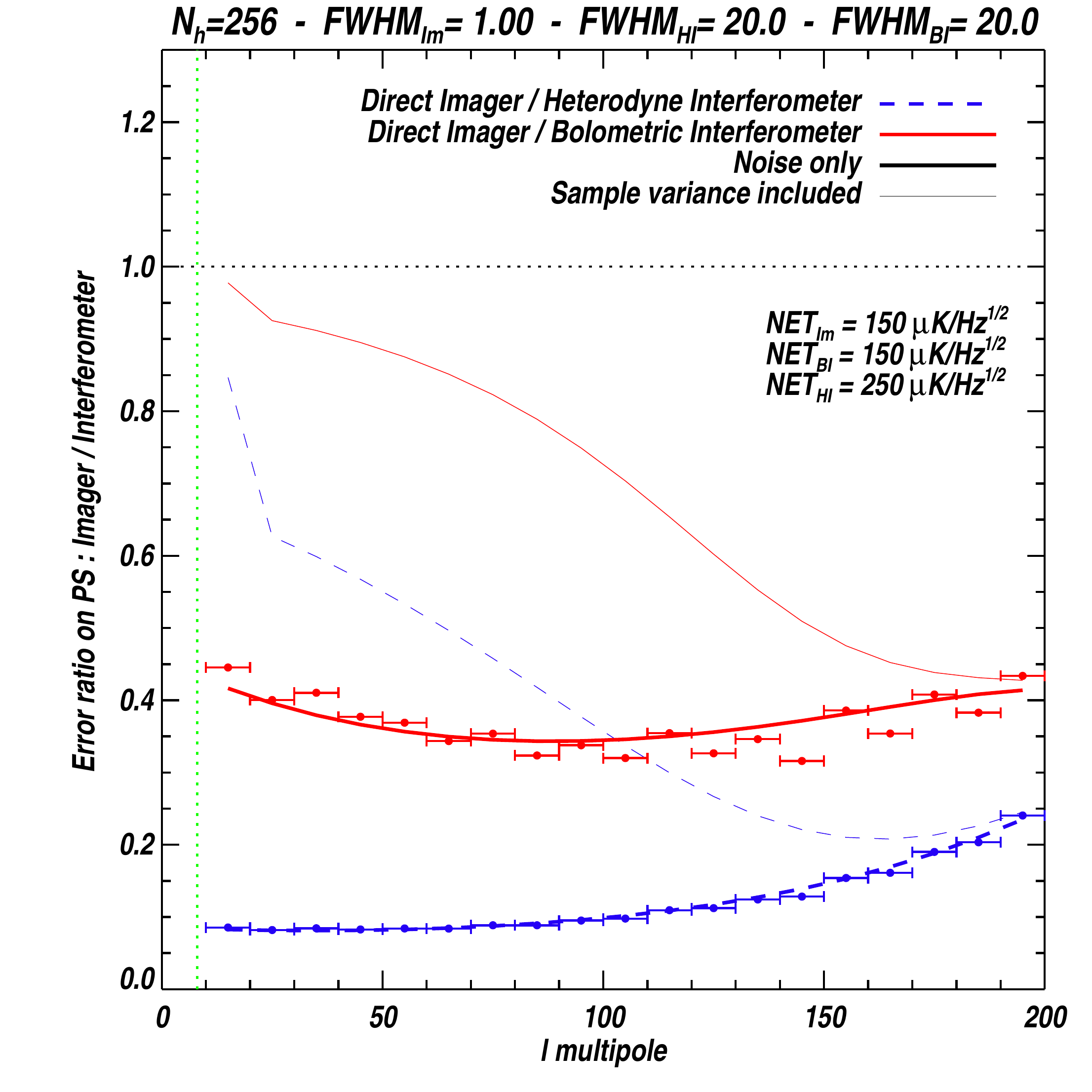}}}
\caption{\small Power spectrum error bar ratio between an imager and a bolometric interferometer (red solid) and between an imager and a heterodyne interferometer (blue dashed). The thick lines only include the noise contribution while the thin lines also include sample variance (with a tensor to scalar ratio of 0.1). In each case we have assumed 256 horns, a sky fraction defined by the 20 degrees FWHM primary beam of the interferometers. The left panel corresponds to a 10 arcminute resolution imager and the right one to a one degree imager. The points are obtained with a precise measurement of the number of equivalent baselines and the lines are from the approximate expression given above. The green dotted line shows the region where the relative sample variance $\Delta C_\ell / C_\ell=\sqrt{2/(2\ell f_\mathrm{sky} \Delta\ell)}$ is greater than one. The ratio of the heterodyne interferometer to the imager depends on the square of the NET ratio; we have assumed 250 and 150 $\mathrm{mK}/ \sqrt{\mathrm{Hz}}$ taken from (\cite{quiet}) and~Clover~(\cite{clover2}) at $\sim $90 GHz for Chile. It is straightforward to scale the curves for different values of the NET. }
\label{comparison} 
\end{figure*}

The number of equivalent baselines for a square horn array is:
\begin{eqnarray}
N_\mathrm{eq}&=&\left(\sqrt{N_h}-\left|\frac{\mathrm{u}_x}{\mathrm{u_{min}}}\right|\right)\left(\sqrt{N_h}-\left|\frac{\mathrm{u}_y}{\mathrm{u_{min}}}\right|\right).
\end{eqnarray}
If one averages over directions in the baseline plane at a given $|\vec{u}|$, a good approximation of $N_\mathrm{eq}$ as a function of $\ell$ is given by (see Fig.~\ref{uapprox}):
\begin{figure}[!t]
\centering\resizebox{\hsize}{!}{\centering{
\includegraphics{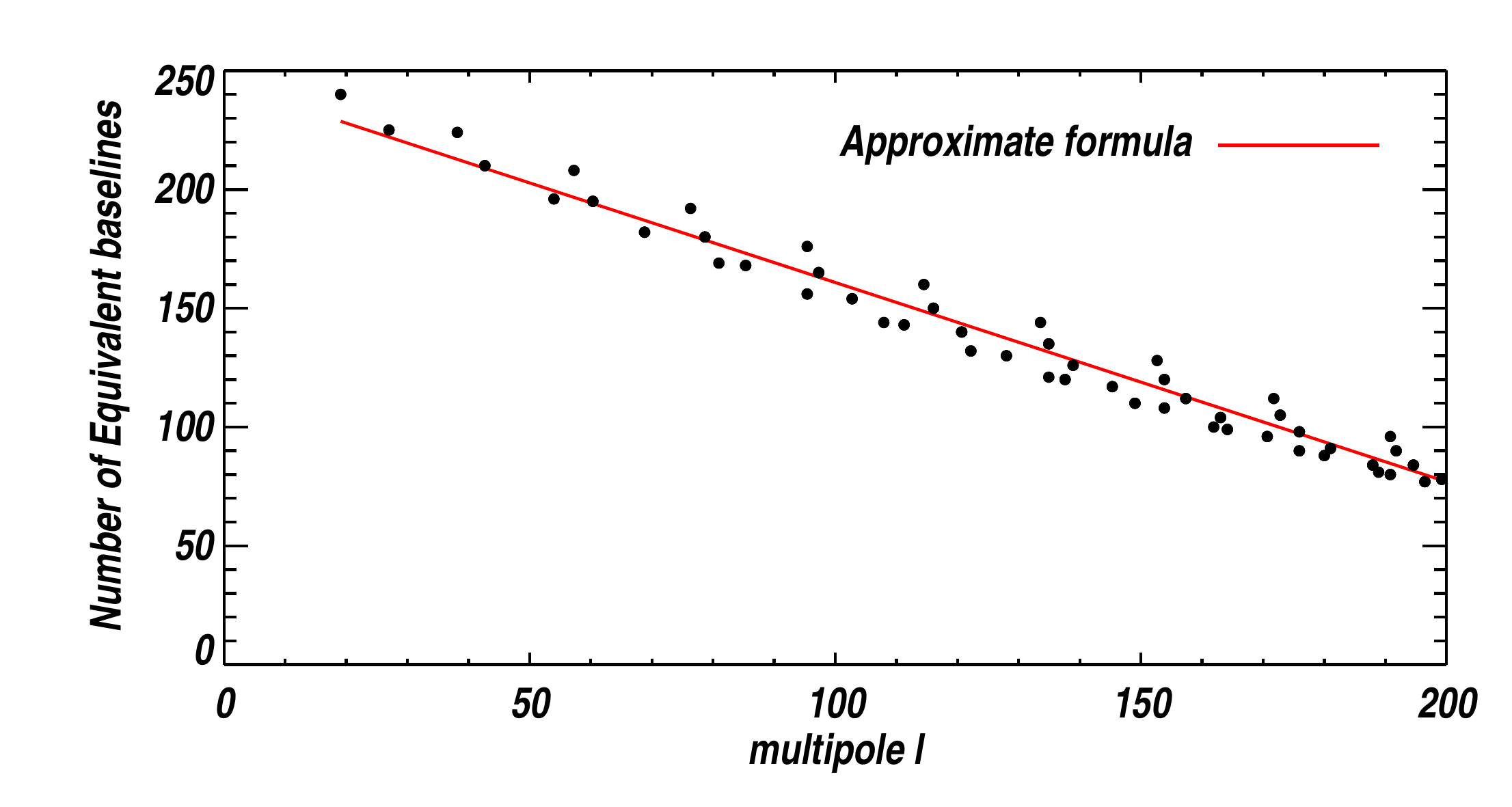}}}
\caption{\small Approximate formula (red line)  from Eq.~\ref{eqapprox} and actual values (black dots) for the number of equivalent baselines as a function of $\ell=2\pi\mathrm{u}$ for 256 entry horns on a square compact grid.}
\label{uapprox} 
\end{figure}
\begin{eqnarray}\label{eqapprox}
\frac{N_\mathrm{eq}}{N_h}\simeq1-\frac{\sqrt{2}}{2\sqrt{N_h}}-\frac{\ell}{\ell_0}~~~~\mathrm{where}~~~\ell_0=\frac{2\sqrt{N_h}}{\sqrt{f_\mathrm{sky}}}.
\end{eqnarray}
One finally finds that a good approximation of the sensitivity ratio is:
\begin{equation}
\left. \frac{\Delta C_\ell^\mathrm{Im}}{\Delta C_\ell^\mathrm{BI}} \right|_\mathrm{noise}\simeq \left(1-\frac{\sqrt{2}}{2\sqrt{N_h}}-\frac{\ell}{\ell_0}\right)^2\times\frac{1}{2~B_\ell^2}
\end{equation}
and:
\begin{equation}
\left. \frac{\Delta C_\ell^\mathrm{Im}}{\Delta C_\ell^\mathrm{HI}}\right|_\mathrm{noise} \simeq \left(1-\frac{\sqrt{2}}{2\sqrt{N_h}}-\frac{\ell}{\ell_0}\right)\times\frac{1}{4~B_\ell^2}\times\left(\frac{\mathrm{NET}_\mathrm{Im}}{\mathrm{NET}_\mathrm{HI}}\right)^2.
\end{equation}
These approximate formulae have been compared with actual calculations of the number of equivalent baselines for square arrays. We have chosen 256 horns for the comparison and we compare bolometric and heterodyne interferometers with imagers having a low angular resolution of one degree, BICEP-like~(\cite{bicep}) and a high one of 10 arcminutes, Clover-like~(\cite{clover}). The results are shown in Fig.~\ref{comparison}. We have chosen to only consider the multipole region between 0 and 200 as for higher multipoles, interferometers are less sensitive due the loss of coherence between largely separated horns. Note that the effect of coherence loss for the long baselines and the bandwidth smearing have not been taken into account here and might have a significant effect.

\section{Discussion\label{discussion}}
The sensitivities of the three different techniques only differ in the way the instrument filters the multipoles observed in the sky. An imager is affected by its resolution on the sky while an interferometer is affected by the ratio between the number of equivalent baselines and the number of horns as a function of multipoles. All of these filtering factors are less than one. However, imagers are usually operated in such a way that they are not limited by their angular resolution in the multipole region of interest, in that case $B_\ell\simeq 1$, and the imager is always more sensitive than an interferometer (bolometric or heterodyne). From the strict point of view of sensitivity, interferometers can therefore only compete with low angular resolution imagers. 

There is a large difference in sensitivity between bolometric and heterodyne interferometers compared to an imager: the ratio $N_\mathrm{eq}/N_h$ acts quadratically on the variance for a bolometric interferometer while it acts linearly for a heterodyne instrument. 
This is due to the fact that with a heterodyne interferometer, equivalent baselines are averaged after their measurement, resulting in a $1/N_\mathrm{eq}$ factor on the variances. In a bolometric interferometer, the signals from all $N_h$ horns are added together multiplying the noise variance by $N_h$ while the coherent summation of equivalent baselines performs an efficient $1/N_\mathrm{eq}^2$ reduction of the noise. This finally results in a factor $N_h/2N_\mathrm{eq}$ for the variance of a bolometric interferometer relative to a heterodyne one.
This is largely compensated by the difference in  NET between bolometric instruments and coherent ones. When comparing them, the ratio of their NET also appears quadratically and favours bolometric instruments that are dominated by the photon noise rather than by that of the amplifiers. This situation may change in the future with the improvements of the HEMT technologies but at frequencies around and above 100 GHz we are unlikely to face photon noise limited HEMTs in the near future. The difference between the NET would be even greater in space where the bolometers NET would drop as the background temperature while that of the coherent instruments would remain roughly constant.

With the present technologies of bolometers and coherent amplifiers, the hierarchy in terms of sensitivity between the three techniques (and layout) studied here is very clear for the multipole range $25<\ell<200$ where the primordial B-mode signal is expected to be maximal. Imagers are the most sensitive, bolometric interferometers have a lower sensitivity, the ratio dropping quadratically with the multipole considered. Heterodyne interferometers have an even lower sensitivity but the ratio with an imager drops less rapidly. They remain however less sensitive than bolometric interferometers in the range of multipoles considered here, where the largest primordial B-mode signal is expected and where the lensing of the E-modes into B-modes is still low. At higher multipoles however, the bandwidth smearing effect and loss of coherence would be a real issue for bolometric interferometers while in a heterodyne interferometer, the separation into small bands would prevent the sensitivity from dropping. 

The main remaining question is whether the gain in terms of systematic effects is worth the price of this sensitivity reduction if one builds an interferometer instead of an imager. 
In terms of optics for instance, an interferometer directly observes the sky. The primary beam is therefore only set by that of the horns, while in an imager, the telescope (mirror or lenses) produces sidelobes inducing poorly predictible  ground pickup that often prevent one from reaching the nominal sensitivity. 
An interferometer is also completely insensitive to spatially uniform polarized signals that vary with time such as polarized atmospheric contamination. These could also prevent an imager from reaching its nominal sensitivity by adding some spread in the noise. These examples mitigate the statistical sensitivity loss of an interferometer with respect to an imager.
The differences in terms of systematic effects between imagers and bolometric and heterodyne interferometers are not obvious and deserve a detailed quantitative study in continuation of the work done by~(\cite{bunn}).

\begin{acknowledgements}
The authors thank Ken Ganga for fruitful discussions and all the participants of the "Bolometric Interferometry for the B-mode search" workshop held in Paris in June 2008 for the many stimulating discussions that took place. We also thank Sarah Church, Keith Grainge and Mike Jones for their explanations concerning heterodyne interferometry.
\end{acknowledgements}

\end{document}